\documentclass[multphys,vecphys]{svmult}
\usepackage{amsfonts}
\usepackage{amssymb}

\usepackage{makeidx}         
\usepackage{graphicx}        
\usepackage{multicol}        
\usepackage[bottom]{footmisc}

\makeindex             
\usepackage{url}

\begin{document}


\title*{Power Spectrum Estimation I. Basics}
\author{Andrew J S Hamilton}
\institute{JILA and Dept.\ Astrophysical \& Planetary Sciences,
Box 440, U. Colorado, Boulder, CO 80309, USA.
\texttt{Andrew.Hamilton@colorado.edu \url{http://casa.colorado.edu/~ajsh/}}}

\svitemindent=16pt


\newcommand{\dd}{\D}
\newcommand{\ddd}{\dd^3}
\newcommand{\e}{\E}
\newcommand{\im}{\I}
\newcommand{\transpose}{\top}
\newcommand{\qex}{\nopagebreak[4]$\qed$\pagebreak[2]}

\newcommand{\el}{\ell}
\newcommand{\nbar}{{\bar n}}
\newcommand{\deltatilde}{{\tilde\delta}}
\newcommand{\ntilde}{{\widetilde n}}
\newcommand{\Ptilde}{{\widetilde P}}

\newcommand{\kvec}{{\vec{k}}}
\newcommand{\Lvec}{{\vec{L}}}
\newcommand{\nvec}{{\vec{n}}}
\newcommand{\pvec}{{\vec{p}}}
\newcommand{\qvec}{{\vec{q}}}
\newcommand{\rvec}{{\vec{r}}}
\newcommand{\xvec}{{\vec{x}}}
\newcommand{\yvec}{{\vec{y}}}
\newcommand{\zvec}{{\vec{z}}}
\newcommand{\onevec}{{\vec{1}}}

\newcommand{\khat}{{\hat\kvec}}
\newcommand{\rhat}{{\hat\rvec}}
\newcommand{\zhat}{{\hat\zvec}}

\newcommand{\mangle}{{\sc mangle}}

\newcommand{\skipp}{\vskip.25cm\noindent{}}

\hyphenpenalty=3000

\maketitle

\section*{Abstract}
This paper and its companion
form an extended version of notes provided to participants
in the Valencia September 2004 summer school on Data Analysis in Cosmology.
The papers offer a pedagogical introduction
to the problem of estimating the power spectrum from galaxy surveys.
The intention is to focus on concepts rather than on technical detail,
but enough mathematics is provided to point the student in the right direction.

This first paper
presents background material.
It collects some essential definitions,
discusses traditional methods for measuring power,
notably the Feldman-Kaiser-Peacock (1994) \cite{FKP94} method,
and introduces Bayesian analysis,
Fisher matrices, and maximum likelihood.
For pedagogy and brevity,
several derivations are set as exercises for the reader.
At the summer school,
multiple choice questions, included herein,
were used to convey some didactic ideas,
and provoked a little lively debate.


\section{Introduction}

It was a flawlessly organised September summer school
in the historic Mediterranean city of Valencia,
whose narrow, marble-paved streets are so randomly variable
that you got lost in them as easily as in one of the lectures
on ``Data Analysis in Cosmology''
going on at the Palacio Pineda.

The lecture on power estimation was one of the first lectures
at the summer school,
and it seemed sensible to make available to the students in advance
a reference set of notes
containing essential definitions and background material
that would prove useful throughout the summer school.
The present paper is a somewhat extended version of
those notes.
The background material in the notes
was not presented at the lecture,
but rather was left as homework for the student during the long hours of siesta.
To facilitate self-study,
several of the derivations are posed as exercises for the reader.
Solutions are not included,
but the derivations contain enough guidance that the persistent
student should be able to solve them.

The power spectrum is the most important statistic that
can be measured from large scale structure (LSS).
During the lecture,
the reasons for this being so
were conveyed through the device of multiple choice questions,
which are included in this paper,
along with answers at the end of the paper.

This paper is arranged as follows.
Section~\ref{definition} collects some essential definitions
of ccrrelation functions, power spectra, and shot noise.
Section~\ref{trad} discusses traditional methods for measuring power,
notably the Feldman-Kaiser-Peacock (1994) \cite{FKP94} method.
Section~\ref{maxlikmethod} introduces Bayesian analysis,
Fisher matrices, and maximum likelihood.

A separate companion paper
focusses on the actual designated topic of the lecture.
It covers the practical issues of
measuring power spectra from observations,
with an emphasis on using maximum likelihood techniques
to measure power at large, linear scales.

\section{Definitions}
\label{definition}

This section collects definitions
of some of the jargon that you will encounter not only in this lecture
but repeatedly throughout this summer school.
It is a good idea to assimilate the jargon\footnote{
I've added some optional footnotes, like this, on Hilbert space.
A Hilbert space
is an infinite dimensional vector space equipped with an inner product.
Hilbert space provides a compact, powerful, and unifying mathematical formalism,
just as ordinary vectors do in finite-dimensional geometry.
A density field is a vector in a Hilbert space;
a covariance function is a matrix in Hilbert space;
an $n$-point correlation function is a rank-$n$ tensor in Hilbert space.
}.

\subsection{Correlation Function}

Let $n(\rvec)$ denote the observed number
\textbf{density}
of particles (galaxies)
at position $\rvec$ in a survey.

Let $\nbar(\rvec)$
denote the
\textbf{selection function},
the expected mean number of particles (galaxies)
at position $\rvec$
given the selection criteria of the survey.
Often but not always,
the selection function is separable into a product of an
\textbf{angular selection function} $\nbar(\rhat)$
and a
\textbf{radial selection function} $\nbar(r)$.
The determination or measurement of the angular and radial selection functions
of a survey is a non-trivial enterprise
which is an essential prerequisite for measuring correlation functions
or power spectra.

The
\textbf{overdensity} $\delta(\rvec)$
of particles (galaxies) at position $\rvec$ is defined by
\begin{equation}
  \delta(\rvec)
  \equiv
  {n(\rvec) - \nbar(\rvec) \over \nbar(\rvec)}
  \;.
\end{equation}

The
\textbf{correlation function} $\xi(r_{ij})$
(or $2$-point correlation function)
is the covariance of overdensities
at separation $r_{ij} \equiv \left| \rvec_i{-}\rvec_j \right|$
\begin{equation}
\label{xi}
  \xi(r_{ij})
  \equiv
  \langle \delta(\rvec_i) \delta(\rvec_j) \rangle
  \;.
\end{equation}
In large scale structure (LSS),
the correlation function is often, though not always,
conventionally taken to refer to
the covariance function in real space
(as opposed to Fourier space or some other space).
The assumption that the Universe is
\textbf{statistically homogeneous}
(= statistically translation invariant)
means that the correlation function is a function
only of the vector separation $\rvec_{ij} \equiv \rvec_i{-}\rvec_j$
of two points.
The assumption that the Universe is
\textbf{statistically isotropic}
(= statistically rotation invariant)
means that the correlation function is a function
only of the magnitude of the separation
$r_{ij} \equiv \left| \rvec_{ij} \right|$
of two points.

\subsection{Power Spectrum}

A
\textbf{Fourier mode} $\delta(\kvec)$
is the Fourier transform of the overdensity\footnote{
You may not be familiar with the practice of using the same symbol
$\delta$ for both real and Fourier space;
but $\delta$ is the same vector in Hilbert space,
with components
$\delta_\rvec$ in real space,
or $\delta_\kvec$ in Fourier space.
The essential property of Hilbert space is the existence of an inner product
(= scalar product).
In the present case, the inner product of two real-valued vectors
$a_\rvec$ and $b_\rvec$
is defined to be
\begin{equation}
\label{ipr}
  a \cdot b
  =
  a^\rvec b_\rvec
  =
  \int a(\rvec) b(\rvec) \, \ddd r
  \;.
\end{equation}
In the index notation
$a^\rvec b_\rvec$,
repeated indices imply implicit summation
(which becomes integration over the infinite dimensional space of positions),
just as in relativity and quantum mechanics.
In repeated pairs of indices,
one index is always raised, while the other is always lowered
(though it is also common, for notational simplicity,
to keep all indices lowered,
which causes no ambiguity as long as it is implicitly understood
that in contracting over paired indices,
one index is always raised and the other lowered).
In real space,
the raised components of a real-valued vector
are numerically equal to the lowered components,
$a^\rvec = a_\rvec$,
but this is a special feature of real space,
and is not true in Fourier space, spherical harmonic space,
or other spaces.
\begin{exercise}
Show from
equation~(\protect\ref{ipr})
and the definition
$a(\kvec) \equiv \int a(\rvec) \e^{\im \kvec . \rvec} \, \ddd r$
of the Fourier transform
that the inner product of vectors
$a_\kvec$ and $b_\kvec$
in Fourier space is
\begin{equation}
  a \cdot b
  =
  a^\kvec b_\kvec
  =
  \int a(\kvec)^\ast b(\kvec) \, {\ddd k \over (2\upi)^3}
\end{equation}
which is called \textbf{Parseval's theorem}.
Once you've set up the formalism,
you can deduce by inspection that
$a^\kvec b_\kvec = a^\rvec b_\rvec$,
since the inner product is by construction a scalar,
independent of the representation of the vectors.
Notice that in Fourier space,
the raised components of a vector
are equal to the complex conjugate of its lowered components,
$a^\kvec = (a_\kvec)^\ast = a_{-\kvec}$.
\qex
\end{exercise}
}
\begin{equation}
\label{deltak}
  \delta(\kvec)
  \equiv
  \int \delta(\rvec) \e^{\im \kvec . \rvec} \, \ddd r
  \;, \quad
  \delta(\rvec)
  =
  \int \delta(\kvec) \e^{- \im \kvec . \rvec} \, {\ddd k \over (2\upi)^3}
  \;.
\end{equation}
The allocation of factors of $2\upi$ here
follows the standard convention in cosmology,
which you would be wise to stick to even if you don't like it.
Other disciplines have their own conventions.

The \textbf{power spectrum} $P(k)$ is the Fourier transform
of the correlation function
\begin{equation}
\label{P}
  P(k)
  \equiv
  \int \xi(r) \e^{\im \kvec . \rvec} \, \ddd r
  \;, \quad
  \xi(r)
  =
  \int P(k) \e^{- \im \kvec . \rvec} \, {\ddd k \over (2\upi)^3}
  \;.
\end{equation}

\begin{exercise}
From the definitions~(\ref{xi}) of the correlation function
and (\ref{deltak}) of Fourier modes,
and the relation~(\ref{P})
between the power spectrum and the correlation function,
show that the covariance of Fourier modes is\footnote{
\addtocounter{exercise}{1}
{\bf Exercise \arabic{exercise}.}
Show that the quantity
$(2\upi)^3 \delta_D(\kvec_i{+}\kvec_j)$
in equation~(\protect\ref{dkdk})
is just the unit matrix
$\onevec_{\kvec_i}^{\kvec_j}$
in Hilbert space.
In other words, show that
$\onevec_{\kvec_i}^{\kvec_j} a_{\kvec_j} = a_{\kvec_i}$
for any vector $a_{\kvec}$.
\qex
}
\begin{equation}
\label{dkdk}
  \langle \delta(\kvec_i) \delta(\kvec_j) \rangle
  =
  (2\upi)^3 \delta_D(\kvec_i + \kvec_j) P(k_i)
\end{equation}
where $\delta_D$ denotes the (here 3-dimensional) Dirac delta-function.
Show that the delta-function arises from the assumption of
statistical translation invariance.
Show that the fact that the power spectrum $P(k_i)$
is a function of the magnitude $k_i \equiv \left| \kvec_i \right|$
of its argument follows from the assumption of
statistical rotation invariance.
\qex
\end{exercise}

\subsection{2-Point Function}

The correlation function or power spectrum are
both representations,
expressed respectively in real space and Fourier space,
of the \textbf{covariance function},
also known as the \textbf{2-point function}.

The 2-point function is the 2nd member of an infinite sequence of
\textbf{$n$-point functions},
which are proportional to the
\textbf{$n$'th order irreducible moments}.
The first irreducible moment is the \textbf{mean}.
The key property of the irreducible moments
is that they are
\textbf{additive over sums of independent density fields}.

\subsection{Shot Noise}

Typically, a galaxy survey samples only some fraction of the galaxies
present in any volume element of the survey.
To proceed, one makes the assumption that
the galaxies surveyed are selected randomly from some continuous
underlying population.

\begin{exercise}
Convince yourself of the theorem that:
\textbf{The correlation function $\xi(r)$ of a discrete random sampling
of a density field is equal to the correlation function of the original field}.
\qex
\end{exercise}

Actually, there's a catch to the above theorem,
which is that the correlation function of the randomly subsampled field
is equal to that of the parent field at all separations
\textbf{except at zero separation}.
If it is allowed that a particle is considered to be a neighbour of itself,
then the correlation function of the randomly subsampled field
acquires an extra contribution, a delta-function at zero separation,
which is the shot noise.

As a general definition,
the
\textbf{shot noise
is the self-particle contribution to any statistic}.
In the case of the correlation function or power spectrum,
the shot noise is the self-pair contribution,
that is, the contribution from pairs consisting of a particle (galaxy)
and itself.

\begin{exercise}
Argue that the shot noise contribution to the correlation function
at a point where the selection function is $\nbar(\rvec)$ is
\begin{equation}
\label{rshot}
  \langle \delta(\rvec_i) \delta(\rvec_j) \rangle_{\rm shot}
  =
  {\delta_D(\rvec_{ij}) \over \nbar(\rvec_i)}
  \;.
\end{equation}
\qex
\end{exercise}

You might think that this is trickery.
Can't you just exclude self-pairs and disregard this shot noise nonsense?
The answer is that if you go to another space, such as Fourier space,
then the shot noise shows up in a way that is not so trivial to remove.

\begin{exercise}
Show from equation~(\ref{rshot}) that the shot noise contribution to the
covariance of Fourier modes is
\begin{equation}
\label{dkdkshot}
  \langle \delta(\kvec_i) \delta(\kvec_j) \rangle_{\rm shot}
  =
  (1 / \nbar) (\kvec_i + \kvec_j)
\end{equation}
where $(1 / \nbar) (\kvec)$ is the Fourier transform of $1 / \nbar(\rvec)$
\begin{equation}
\label{nbark}
  (1 / \nbar) (\kvec)
  \equiv
  \int \left[ 1 / \nbar(\rvec) \right] \e^{\im \kvec . \rvec} \, \ddd r
  \;.
\end{equation}
\qex
\end{exercise}

According to equations~(\ref{dkdkshot}) and (\ref{nbark}),
the shot noise contribution to the variance of Fourier modes is
$\langle \delta(\kvec) \delta(\kvec)^\ast \rangle_{\rm shot} = (1 / \nbar) (0)$.
For any finite survey,
this shot noise contribution is infinite.
This simply reflects the fact that
Fourier modes are waves extending to infinity,
and that it would require an infinite survey to measure the
the amplitude of a wave whose wavenumber is specified with infinite precision.
Fourier modes of real finite surveys
are subject to an uncertainty principle:
the wavenumbers of their Fourier modes are not precise,
but rather are smeared over some finite width $\varDelta k \sim 1 / R$,
where $R$ is a measure of the linear size of the survey.
You will discover more about what happens in real surveys
in the exercise in \S\ref{BF}.

\pagebreak[2]
\begin{question}
\label{choice}
\textbf{Why is the 2-point function
(either the correlation function or the power spectrum)
the statistic of choice in characterizing LSS?}
All of the following are true,
but which is the most important?
\begin{itemize}
\item[A.]
Because it has a simple physical meaning:
the correlation function $\xi(r)$
is the average excess over random
of the probability of finding a
particle (galaxy) at given separation $r$
from another particle (galaxy).
\item[B.]
Because the 2-point function can be measured relatively
easily from observations, essentially by counting pairs.
\item[C.]
Because the correlation function or power spectrum
is the (co)variance of density (the 2nd irreducible moment),
which is the lowest order irreducible moment
after the mean (the 1st irreducible moment).
\item[D.]
Because the central limit theorem
implies that
a density distribution is asymptotically Gaussian
in the limit where the density results from
the average of many independent random processes;
and a Gaussian is completely characterized by
its mean and variance
(the 1st and 2nd irreducible moments).
\item[E.]
Because the 2-point function satisfies a dynamical equation,
a low order member of the BBGKY hierarchy of equations.
\end{itemize}

Answer at end of paper.
\qex
\end{question}

\begin{question}
\label{goodpk}
\textbf{
What is the advantage of the power spectrum $P(k)$
over the correlation function $\xi(r)$?}
Which of the following is the most important?
\begin{itemize}
\item[A.]
During the linear growth of fluctuations,
the evolution of the Fourier mode $\delta(\kvec)$
at each wavevector $\kvec$ is independent of every other.
\item[B.]
The
covariance matrix of Fourier modes $\delta(\kvec)$
is a diagonal matrix, equation~(\ref{dkdk}),
whereas the covariance matrix of real space modes $\delta(\rvec)$
is {\em not\/} a diagonal matrix, equation~(\ref{xi}).
\item[C.]
The power spectrum is the covariance of Fourier modes;
Fourier modes are eigenmodes of the translation operator;
the density distribution is statistically translation invariant;
hence the cosmic covariance matrix must commute with the translation operator.
\item[D.]
Estimates of the power $P(k)$
at different wavenumbers $k$ are uncorrelated,
for Gaussian fluctuations,
whereas estimates of the correlation function $\xi(r)$
at different separations $r$ are correlated.
\item[E.]
The power spectrum is easier to measure than the correlation function.
\end{itemize}

Answer at end of paper.
\qex
\end{question}

\pagebreak[4]
\section{Traditional Methods for Measuring Power}
\label{trad}

Yu \& Peebles (1969) \cite{YP69} and Peebles (1973) \cite{P73}
were the first to characterize LSS with the power spectrum.
Their methodology was complicated by the fact that they had only
positions on the sky, not full 3-dimensional positions.

Baumgart \& Fry (1991) \cite{BF91} first pointed out
that you could measure the galaxy power spectrum $P(k)$ in a redshift survey
by the simple method of enclosing the survey in a box and Fourier transforming
\textbf{without having to bother about the detailed boundaries of the survey}.
This astonishing result appeared to be in stark contrast to
measurements of the correlation function $\xi(r)$,
where it was essential to worry about the survey boundaries.

In an influential paper,
Feldman, Kaiser \& Peacock (1994) \cite{FKP94}
proposed a variant of the Baumgart \& Fry \cite{BF91} method,
in which each galaxy $i$ is first weighted by
\begin{equation}
\label{FKPweight}
  w_i = {1 \over 1 + \nbar(\rvec_i) P(k)}
\end{equation}
where $\nbar(\rvec_i)$ is the selection function at the position
$\rvec_i$ of galaxy $i$,
before Fourier transforming.
FKP showed that this procedure provided an optimal estimate of power $P(k)$
in the case that
\\
(1) the wavelength $2\upi / k$ is small compared to the scale of the survey,
and
\\
(2) fluctuations are Gaussian.
\\
Physically,
the FKP weighting~(\ref{FKPweight})
is an approximation to the inverse variance weighting.
It weights volumes by $1 / \left[ \nbar^{-1}(\rvec) + P (k) \right]$,
which one recognizes as the reciprocal of the sum of the shot
noise $\nbar^{-1}(\rvec)$ and the cosmic power $P(\kvec)$.
This approximation is valid only in the ``classical'' limit
where position $\rvec$ and wavenumber $\kvec$ are simultaneously measurable,
which is condition (1) above.
The condition (2) of Gaussianity
comes from the fact that the thing being measured is power,
a 2-point statistic,
and the uncertainty in power involves,
in addition to a product of 2-point terms,
a 4-point term which vanishes only for Gaussian fluctuations.

Notice that the FKP weighting~(\ref{FKPweight})
depends on the power spectrum $P(k)$,
which is the same thing that the FKP method aims to measure.
In Bayesian analysis
the $P(k)$ in the FKP weight would be recognized as being part of the model
(part of the prior).
However, the FKP approach is not Bayesian,
but rather follows traditional statistical methods.

The FKP method is excellent for the intuition,
and for quick, approximate estimates.
However, it is not adequate for precision cosmology
and the estimation of cosmological parameters.
The FKP method is inadequate both at large scales
where assumption (1) fails,
and at small scales,
where assumption (2) fails.
The problem is not so much that the FKP weighting is suboptimal
(though that is true),
but rather that the FKP method does not yield a precise estimate
of the variance and covariance of measured power.
Moreover, it is not powerful enough to deal with all of the real world issues
of actual galaxy surveys.

\subsection{The Baumgart \& Fry (1991) \cite{BF91} Miracle.}
\label{BF}

\begin{exercise}
\vskip.05cm
Let $\deltatilde(\rvec)$
denoted a weighted overdensity of galaxies
at position $\rvec$ in a survey:
\begin{equation}
\label{deltatilde}
  \deltatilde(\rvec)
  \equiv w(\rvec) \left[ n(\rvec) - \nbar(\rvec) \right]
  = w(\rvec) \nbar(\rvec) \delta(\rvec)
  = W(\rvec) \delta(\rvec)
  \;.
\end{equation}
Here $w(\rvec)$ is some arbitrary weighting (such as the FKP weighting)
that you choose
[with the proviso that the weighting must be chosen a priori,
independent of the observed galaxy density $n(\rvec)$];
and $W(\rvec) \equiv w(\rvec) \nbar(\rvec)$
is the product of the weighting function and the selection function.

\begin{itemize}
\item[(a)]
\textbf{Fourier modes of the weighted overdensity}
\vskip.05cm
The Fourier transform of the weighted overdensity $\deltatilde(\rvec)$ is,
by definition,
\begin{equation}
  \deltatilde(\kvec)
  = \int \deltatilde(\rvec) \e^{\im \kvec . \rvec} \, \ddd r
  = \int W(\rvec) \delta(\rvec) \e^{\im \kvec . \rvec} \, \ddd r
  \;.
\end{equation}
Show that $\deltatilde(\kvec)$
equals the convolution of the Fourier transform $W(\kvec)$
of the window
with the Fourier transform $\delta(\kvec)$ of the overdensity:
\begin{equation}
  \deltatilde(\kvec)
  =
  \int W(\kvec - \kvec^\prime) \delta(\kvec^\prime)
  \, {\ddd k \over (2\upi)^3}
\end{equation}
This is just the standard result that multiplication in real space
becomes convolution in Fourier space.

\item[(b)]
\textbf{Covariance of Fourier modes of the weighted overdensity}
\vskip.05cm
Assume that the covariance
$\langle \delta(\kvec_i) \delta(\kvec_j) \rangle$
of (unweighted) overdensities in Fourier space
is a sum of a cosmic term
$(2\upi)^3 \delta_D(\kvec_i{+}\kvec_j) P(k_i)$,
equation~(\ref{dkdk}),
and a shot noise term
$(1 / \nbar) (\kvec_i{+}\kvec_j)$,
equation~(\ref{dkdkshot}),
\begin{equation}
\label{didj}
  \langle \delta(\kvec_i) \delta(\kvec_j) \rangle
  =
  (2\upi)^3 \delta_D(\kvec_i + \kvec_j) P(k_i)
  +
  (1 / \nbar) (\kvec_i + \kvec_j)
  \;.
\end{equation}
Show that the covariance of Fourier modes of the weighted overdensity is
\begin{equation}
  \langle \deltatilde(\kvec_i) \deltatilde(\kvec_j) \rangle
  =
  \int
  W(\kvec_i - \kvec) W(\kvec_j + \kvec) P(\kvec)
  \, {\ddd k \over (2\upi)^3}
  \ + \ 
  N(\kvec_i + \kvec_j)
\end{equation}
where the shot noise $N(\kvec)$ is\footnote{
Actually
it is more accurate to use the {\bf actual} value of the shot noise,
which is
\begin{equation}
\label{Nactual}
  N(\kvec)
  =
  \sum_{{\rm galaxies} ~ i} w(\rvec_i)^2 \, \e^{\im \kvec . \rvec_i}
\end{equation}
as opposed to equation~(\protect\ref{Nexpect})
which merely gives the {\bf expectation value} of the shot noise.
Shot noise is, by definition,
the contribution to the covariance from self-pairs
(pairs consisting of a particle and itself).
Equation~(\ref{didj}),
from which follows equation~(\ref{Nexpect}),
is a statement about the average excess of neighbours of a particle.
But in fact we know more about the shot noise than just its average:
we know that each particle always has exactly one of itself as a neighbour,
not merely on average.
In statistics,
an estimate that uses more prior information
is always better than an estimate that uses less.
}
\begin{equation}
\label{Nexpect}
  N(\kvec)
  =
  \int {w(\rvec)^2 \nbar(\rvec)} \e^{\im \kvec . \rvec} \, \ddd r
  \;.
\end{equation}
Hence conclude that the variance of Fourier modes of the weighted overdensity is
\begin{equation}
\label{dtkdtk}
  \langle \deltatilde(\kvec) \deltatilde(\kvec)^\ast \rangle
  =
  \int
  \left| W(\kvec - \kvec^\prime) \right|^2 P(\kvec^\prime)
  \, {\ddd k^\prime \over (2\upi)^3}
  \ + \ 
  N(0)
  \;.
\end{equation}
Equation~(\ref{dtkdtk}) says that the variance of Fourier modes
of weighted overdensity is, after subtraction of the shot noise $N(0)$,
equal to the power spectrum $P(\kvec)$ smoothed over
a smoothing function given by the magnitude squared of the
$\left| W(\kvec) \right|^2$
of the Fourier transform $W(\kvec)$ of the window.
Let $W^2$ denote the integral over the window
(a notation suggested by the fact that $W^2$ is
the scalar product of the Hilbert-space vector $W_\kvec$ with itself)
\begin{equation}
  W^2
  =
  \int
  \left| W(\kvec) \right|^2
  \, {\ddd k \over (2\upi)^3}
  =
  \int W(\rvec)^2 \, \ddd r
  \;.
\end{equation}
Then a smoothed estimate $\Ptilde(\kvec)$ of power at wavevector $\kvec$ is
\begin{equation}
\label{Ptilde}
  \Ptilde(\kvec)
  \equiv
  {\int
  \left| W(\kvec - \kvec^\prime) \right|^2 P(\kvec^\prime)
  \, \ddd k^\prime / (2\upi)^3
  \over
  W^2}
  =
  {\langle \deltatilde(\kvec) \deltatilde(\kvec)^\ast \rangle - N(0)
  \over
  W^2}
  \;.
\end{equation}
Equation~(\ref{Ptilde}) is essentially
Baumgart \& Fry's (1991) \cite{BF91} remarkable result.

\item[(c)]
\textbf{What does it mean?}
\vskip.05cm
Suppose that the survey window $W(\rvec)$ has a characteristic size $R$.
Approximately what is the width of the smoothing window
$\left| W(\kvec) \right|^2$
in the smoothed power spectrum, equation~(\ref{Ptilde})?
At what wavenumber $k$ would you cease to trust
the smoothed power spectrum $\Ptilde(k)$
as a reasonable estimate of the true power $P(k)$?
What happens as the size $R$ of the survey gets larger?
[The important thing here is the concept rather than the mathematics.
But if you want to see how this plays out mathematically,
you might consider a survey window $W(\rvec)$ which
happens to be a Gaussian
$W(\rvec) = \exp\left[- r^2 / (2 R^2)\right]$,
centred at the origin, with a 1-$\sigma$ width of $R$.]
\end{itemize}
\qex
\end{exercise}

\subsection{The Feldman-Kaiser-Peacock (1994) \cite{FKP94} Method}
\label{FKP}

In the previous exercise
you obtained an estimate, equation~(\ref{Ptilde}),
of (smoothed) power $\Ptilde(\kvec)$
at wavevector $\kvec$.
The estimate involved an arbitrarily adjustable weighting $W(\rvec)$,
equation~(\ref{deltatilde}),
of volume elements in the survey.
It is natural to try to choose this weighting $W(\rvec)$ to try to minimize
the variance of the resulting estimate~(\ref{Ptilde}) of power.
The FKP weighting,
already given as equation~(\ref{FKPweight}),
is an approximation to the desired minimum variance weighting,
valid under the two conditions stated
immediately after equation~(\ref{FKPweight}).

It proves surprisingly tricky to derive the FKP weighting in
a rigorous way with a minimum of unnecessary assumptions.
It would be nice to take you through the derivation in an exercise,
but I could not devise an approach that was satisfactorily
clean, insightful, and brief.
You might like to consult the original
Feldman, Kaiser \& Peacock (1994) \cite{FKP94} paper
to see how they did it.
A more general derivation can be found in Hamilton (1997) \cite{H97}.
Perhaps the most elegant approach is to use the quadratic method
of Tegmark (1997) \cite{T97},
discussed in \S12 of Paper~2.

A core part of the FKP argument is the following.
If the survey has characteristic linear size $\sim R$,
then the Fourier transform $W(\kvec)$ of the survey window
will be a ball of width $\varDelta k \sim 1 / R$
around the origin $\kvec = 0$.
It follows that
at wavenumbers much larger than the reciprocal size of the survey,
$k \gg 1 / R$,
the smoothing window
$\left| W(\kvec - \kvec^\prime) \right|^2$
in the Baumgart \& Fry estimate~(\ref{Ptilde})
will be narrowly concentrated around the target wavenumber $\kvec$.
To the extent that the power spectrum $P(\kvec^\prime)$ is slowly
varying over the narrow window,
it can be approximated by a constant,
\begin{equation}
  P(\kvec^\prime) \approx
  P(k) = \mbox{constant}
  \;.
\end{equation}
If the power spectrum is interpreted as literally constant,
then the covariance matrix of overdensities is diagonal in real space
\begin{equation}
\label{drdr}
  \langle \delta(\rvec_i) \delta(\rvec_j) \rangle
  =
  \delta_D \left( \rvec_i - \rvec_j \right)
  \left[ P(k) + {1 \over \bar n( \rvec_i )} \right]
  \ .
\end{equation}
This indicates that each volume element of the survey can be approximated
as being statistically uncorrelated with all other volume elements.
If you buy into the notion
that minimum variance weighting is inverse variance weighting
(\S\ref{fishermx} explains where that notion comes from),
then equation~(\ref{drdr})
suggests that each volume element should be weighted by
\begin{equation}
  W(\rvec) = {1 \over P(k) + 1 / \bar n( \rvec )}
  \ .
\end{equation}
In the present case, the thing of interest is not single volume elements,
but rather pairs of volume elements.
For the specific case of Gaussian fluctuations,
the covariance of pairs
is a product of covariances of singles
(e.g.\ Hamilton 1997 \cite{H97} \S2.1)
\begin{equation}
  \left\langle
    \left(
      \delta_i \delta_j
      - \langle \delta_i \delta_j \rangle
    \right)
    \left(
      \delta_k \delta_l
      - \langle \delta_k \delta_l \rangle
    \right)
  \right\rangle
  =
  \langle \delta_i \delta_k \rangle
  \langle \delta_j \delta_l \rangle
  +
  \langle \delta_i \delta_l \rangle
  \langle \delta_j \delta_k \rangle
\end{equation}
which is true in both real space and Fourier space
(it is a covariant expression in Hilbert space).
It follows that,
for Gaussian fluctuations,
the inverse variance weighting of pairs of volume elements is
\begin{equation}
\label{FKPprweight}
  W(\rvec_i) W(\rvec_j)
  =
  {1 \over
  \left[ P(k) + 1 / \bar n( \rvec_i ) \right]
  \left[ P(k) + 1 / \bar n( \rvec_j ) \right]}
  \ .
\end{equation}
Equation~(\ref{FKPprweight}) is the FKP pair weighting.

\section{Bayes, Fisher, and Maximum Likelihood}
\label{maxlikmethod}

Bayesian statistics provides
the modern mathematical framework for rigorous statistics.
As explained below, \S\ref{maxlik},
it gives special status to maximum likelihood
as yielding the best estimate of a parameter or set of parameters.

Fisher, Scharf \& Lahav (1994) \cite{KSL94} were the first to apply
a likelihood approach to large scale structure.
Heavens \& Taylor (1995) \cite{HT95}
may be credited with accomplishing the first likelihood analysis
designed to retain as much information as possible at linear scales.
With the work of Heavens \& Taylor,
maximum likelihood methods appeared on the LSS scene essentially fully fledged.

\subsection{Bayesian Statistics}
\label{Bayes}

{\bf Traditional statistics.}
Measure the mean by measuring the mean;
measure the variance by measuring the variance;
and such-like na\"{\i}vety.

{\bf Bayesian statistics}.
Measure the mean (or variance) by asking,
what is the probability that the mean (or variance)
takes such and such a value,
given
this set of observations
and
this set of assumptions (the prior).
Commonly,
the prior is subdivided into
(a) assumptions that you assert are true,
and
(b) a model equipped with parameters whose values you wish to estimate.

The foundation of Bayesian statistics foundation is Bayes' theorem.

{\bf Bayes' Theorem}:
States that
the \textbf{posterior probability}
$P( p | x, X )$
that the parameters $p$ take on certain values,
given the observational data $x$ and prior assumptions $X$,
is proportional to
the \textbf{likelihood function},
the probability
$P( x | p, X )$
of the observations $x$
given parameters $p$ and prior assumptions $X$,
multiplied by the \textbf{prior probability}
$P( p | X )$
of the parameters $p$ given the prior assumptions $X$
\begin{equation}
  P( p | x, X )
  =
  {P( x | p, X )
  \,
  P( p | X )
  \over
  P( x | X )}
  \;.
\end{equation}
\qex\\
The probability $P( x | X )$
of the observations given the prior assumptions
is an overall normalization constant
which plays no role except to ensure that
the integral over the posterior probability is 1.

The likelihood function
$P( x | p, X )$,
which converts by multiplication
a prior probability into a posterior probability,
encapsulates all the information provided by a set of observations.
In recognition of this fundamental role,
the likelihood is given its own letter,~${\cal L}$.
R. A. Fisher,
who developed much of the formalism of likelihoods
in the first half of the 20th century,
never subscribed to Bayesian statistics
-- after all, as just remarked,
the likelihood
encapsulates all the information provided by a set of observations.
Nevertheless,
if you want to convert a likelihood into a (posterior) probability
that the parameters take on a certain range of values,
then you have to assume a prior probability distribution of parameters.

``If the prior matters, then you are not learning much from the data''
-- from an Aspen Center for Physics workshop in summer 1997.

\begin{question}
\label{priors}
\textbf{Rank each of the following prior assumptions
in order of probability of being true:}
\begin{itemize}
\item[A.]
The Universe is statistically homogeneous and isotropic.
\item[B.]
The growth of fluctuations is driven primarily by gravity.
\item[C.]
Fluctuations at linear scales are Gaussian.
\item[D.]
The Universe is spatially flat.
\item[E.]
The $\varLambda$CDM model, with
$\varOmega_\varLambda \approx 0.7$,
$\varOmega_c \approx 0.26$,
and $\varOmega_b \approx 0.04$,
is correct
($\varLambda$ is the cosmological constant,
$c$ is Cold Dark Matter,
and $b$ is baryons).
\item[F.]
Galaxy bias $b(k)$,
defined to be the square root of the ratio
of galaxy-galaxy power $P_{gg}(k)$
to matter-matter power $P_{mm}(k)$,
\begin{equation}
  b(k)^2 = {P_{gg}(k) \over P_{mm}(k)}
  \;,
\end{equation}
is constant at linear scales.
\end{itemize}

Answer at end of paper.
\qex
\end{question}

\subsection{Fisher Information Matrix}
\label{fishermx}

The {\bf Fisher information matrix}
(Fisher 1935 \cite{F35})
plays a fundamental role in Bayesian statistical analysis.
Many of us in the field of large scale structure
learned about Fisher matrices from the superb paper by
Tegmark, Taylor \& Heavens (1996) \cite{TTH97},
and I can offer little better advice than to go read that paper!

The term ``optimal'', applied to some statistical estimate of a quantity,
has acquired a bad reputation thanks to misuse.
The Fisher matrix puts what is meant by ``optimal''
on a sound mathematical basis.
It is well worth getting your brain around the Fisher matrix,
because it will raise your understanding of statistics to a new level.

The Fisher information matrix $F_{\alpha\beta}$ of a set of parameters
$p_\alpha$
is formally defined to be minus the expectation value
of the second derivative of the log-likelihood function with respect
to the parameters:
\begin{equation}
\label{Fab}
  F_{\alpha\beta}
  \equiv
  - \left\langle {\partial^2 \ln {\cal L} \over \partial p_\alpha \partial p_\beta} \right\rangle
  \;.
\end{equation}
Expectation value here means
averaged over an ensemble of observational data $x$ predicted by the
likelihood function
\begin{equation}
\label{expect}
  \langle t \rangle \equiv \int t \, {\cal L}( x | p ) \, \dd x
  \;.
\end{equation}
Since the likelihood ${\cal L}$ is multiplicative
over statistically independent sets of observations,
it follows that the Fisher matrix, equation~(\ref{Fab}),
is additive over statistically independent observations,
a sensible property for information to have.

The power of the Fisher matrix derives from the Cram\'{e}r-Rao inequality
(Kendall \& Stuart 1967 \cite{KS67} \S17.15),
which states that the variance
$\left\langle \varDelta \hat p_\alpha ^2 \right\rangle$
of any unbiassed (see equation~(\ref{unbias}) below)
estimate $\hat p_\alpha$ of a parameter $p_\alpha$
must exceed the reciprocal of the diagonal element of the Fisher matrix:
\begin{equation}
\label{CR}
  \left\langle \varDelta \hat p_\alpha ^2 \right\rangle
  \geq {1 \over F_{\alpha\alpha}}
  \quad \mbox{(no summation over $\alpha$)}
  \;.
\end{equation}
You will derive the Cram\'{e}r-Rao inequality
in Exercise~\ref{CramerRao} below.

To the extent that 
the likelihood function
${\cal L}$
is a Gaussian about its maximum
(this is distinct from the proposition that the likelihood
function is Gaussian in the data),
often a good approximation thanks to the central limit theorem,
the Fisher matrix is approximately equal to the inverse
of the covariance matrix of the parameters.
You are probably familiar from your earliest statistical training
with the notion that
the ``best'' way to weight a set of data is by their inverse variance
(an idea already encountered in \S\ref{FKP} on the FKP weighting).
Inverse variance weighting effectively weights data
in proportion to the amount of information in each part.

\subsection{Maximum Likelihood}
\label{maxlik}

An {\bf estimator} $\hat p$ of a parameter $p$
(the hat on the estimator $\hat p$ distinguishes it from the true value $p$)
is some function $\hat p(x)$ of the observational data $x$.
An estimate $\hat p$ of a parameter $p$ is \textbf{unbiassed} if
the expectation value, equation~(\ref{expect}),
of the estimate equals the true value
\begin{equation}
\label{unbias}
  \langle \hat p \rangle = p
  \;.
\end{equation}

A theorem of fundamental importance
(Kendall \& Stuart 1967 \cite{KS67} \S18.5)
states that if an unbiassed estimator attaining the Cram\'{e}r-Rao bound exists,
then it is the {\bf maximum likelihood} estimator,
the values $\hat p_\alpha$ of the parameters
for which the likelihood attains its maximum value
given the observed data:
\begin{equation}
  \left. {\partial \ln {\cal L} \over \partial p_\alpha} \right|_{p_\alpha = \hat p_\alpha} = 0
  \;.
\end{equation}
It is this theorem that gives the maximum likelihood method its special status.

Yet more theorems give conditions
(see Kendall \& Stuart 1967 \cite{KS67},
and Exercise~\ref{CramerRao}(d) below)
under which an unbiassed estimator attaining the Cram\'{e}r-Rao bound exists.
For example,
such an estimator exists if
the likelihood function
${\cal L}$
is a Gaussian about its maximum.
The central limit theorem ensures that
${\cal L}$
is asymptotically Gaussian in the limit of a large amount of data.
Thus
an unbiassed estimator attaining the Cram\'{e}r-Rao bound exists
in the asymptotic limit of a large amount of data.

\begin{exercise}{{\bf The Schwarz inequality.}}
\label{Schwarz}
\vskip.05cm
The basis of the Cram\'{e}r-Rao inequality is the Schwarz inequality,
which states, equation~(\ref{Sch2}) below,
that the correlation coefficient between two estimates
must be less than or equal to one in absolute value.
The Schwarz inequality is a powerful general result in statistics,
and it is well worth knowing how to derive it.

\begin{itemize}
\item[(a)]
Consider two estimators $\hat p$ and $\hat q$,
with (co)variances
$\langle \varDelta \hat p^2 \rangle$,
$\langle \varDelta \hat p \varDelta \hat q \rangle$,
and
$\langle \varDelta \hat q^2 \rangle$.
It is evidently true that
\begin{equation}
\label{Sch1}
  \langle ( \varDelta \hat p + \lambda \varDelta \hat q )^2 \rangle \geq 0
\end{equation}
for any real number $\lambda$.
For what value of $\lambda$, in terms of
the (co)variances,
is the left hand side of equation~(\ref{Sch1}) a minimum?

\item[(b)]
Hence derive the {\bf Schwarz inequality}
\begin{equation}
\label{Sch2}
  \left|
  {\langle \varDelta \hat p \varDelta \hat q \rangle \over
  \langle \varDelta \hat p^2 \rangle^{1/2}
  \langle \varDelta \hat q^2 \rangle^{1/2}}
  \right|
  \leq 1
  \;.
\end{equation}
The quantity inside the vertical bars
on the left hand side of equation~(\ref{Sch2}) is
called the {\bf correlation coefficient} of the estimators $\hat p$ and $\hat q$.
The Schwarz inequality states that correlation coefficient
must lie in the interval $[-1, 1]$.
A correlation coefficient of $1$ means that the estimators are perfectly correlated;
a correlation coefficient of $-1$ means that the estimators are perfectly anti-correlated.

\item[(c)]
What relation between $\hat p$ and $\hat q$ must be satisfied for
the Schwarz inequality to become an equality?
[Answer:
$\varDelta \hat p$ must be proportional to $\varDelta \hat q$.]
\end{itemize}
\qex
\end{exercise}

\begin{question}
\label{Schpk}
{\bf Which of the following is true
according to the Schwarz inequality?}
In the following,
the subscripts ${\rm g}$ and ${\rm m}$ denote galaxies and matter,
so that, for example,
$\xi_{\rm gg}(r)$
and
$\xi_{\rm mm}(r)$
are the galaxy-galaxy and matter-matter correlation functions,
while $\xi_{\rm gm}(r)$ is the galaxy-matter cross correlation.
\begin{itemize}
\item[A.]
$\left| \xi_{\rm gm}(r) / \left[ \xi_{\rm gg}(r) \xi_{\rm mm}(r) \right]^{1/2} \right| \leq 1$.
\item[B.]
$\left| P_{\rm gm}(k) / \left[ P_{\rm gg}(k) P_{\rm mm}(k) \right]^{1/2} \right| \leq 1$,
where $P$ is the shot-noise-subtracted power spectrum.
\item[C.]
$\left| P_{\rm gm}(k) / \left[ P_{\rm gg}(k) P_{\rm mm}(k) \right]^{1/2} \right| \leq 1$,
where $P$ is the power spectrum before shot noise is subtracted.
\item[D.]
All of the above.
\item[E.]
None of the above.
\end{itemize}

Answer at end of paper.
\qex
\end{question}

\begin{exercise}{{\bf The Cram\'{e}r-Rao inequality.}}
\label{CramerRao}
\vskip.05cm
In this exercise you will derive the Cram\'{e}r-Rao inequality,
equation~(\ref{CR}).
The derivation follows
Kendall \& Stuart (1967) \cite{KS67} \S17.15,
which you should consult for more detail and generality.
For simplicity, the exercise considers just a single parameter $p$.

\begin{itemize}
\item[(a)]
The likelihood function ${\cal L}( x | p )$
is the probability of the data $x$ given the parameter $p$, and so
satisfies the normalization condition
\begin{equation}
\label{CR1}
  \int {\cal L}( x | p ) \, \dd x = 1
\end{equation}
for any value of the parameter $p$.
Differentiate this with respect to the parameter $p$ to obtain
$\int \left( \partial {\cal L} / \partial p \right) \, \dd x = 0$,
or equivalently
\begin{equation}
\label{CR2}
  \left\langle {\partial \ln{\cal L} \over \partial p} \right\rangle
  = 0
  \;.
\end{equation}
Differentiate again with respect to $p$ to obtain
\begin{equation}
\label{CR3}
  \mbox{}
  - \left\langle {\partial^2 \ln{\cal L} \over \partial p^2} \right\rangle
  =
  \left\langle \left( {\partial \ln{\cal L} \over \partial p} \right)^2 \right\rangle
  \;.
\end{equation}
You recognise the left hand side of~(\ref{CR3})
as the Fisher information in the parameter $p$,
equation~(\ref{Fab}),
and you see from equation~(\ref{CR3})
that this information must be positive.

\item[(b)]
Consider an unbiassed estimator $\hat p$.
Being unbiassed, equation~(\ref{unbias}),
the estimator must satisfy
\begin{equation}
\label{CR4}
  \left\langle \hat p - p \right\rangle
  =
  \int ( \hat p - p ) {\cal L} \, \dd x
  =
  0
  \;.
\end{equation}
Differentiate this with respect to $p$ to show that
\begin{equation}
\label{CR5}
  \left\langle ( \hat p - p ) {\partial \ln {\cal L} \over \partial p} \right\rangle
  =
  1
  \;.
\end{equation}

\item[(c)]
Apply the Schwarz inequality to deduce from equation~(\ref{CR5}),
coupled with equation~(\ref{CR3}),
the Cram\'{e}r-Rao inequality
\begin{equation}
\label{CR6}
  \left\langle ( \hat p - p )^2 \right\rangle
  \geq
  {1 \over - \left\langle {\partial^2 \ln {\cal L} / \partial p^2} \right\rangle}
  \;.
\end{equation}

\item[(d)]
In question~\ref{Schwarz}(d)
you obtained a condition for the Schwarz inequality to become an equality.
What is the condition on the likelihood function ${\cal L}$
for the Cram\'{e}r-Rao bound to be attained?
[Before you rush to the conclusion that ${\cal L}$ must be Gaussian,
consider (and prove) the fact that the Cram\'{e}r-Rao bound is also
attained by a Poission distribution,
for which the likelihood of observing $x$ counts
over an interval during which
the expected number of counts is $p$ is
$(p)^x \e^{-p} / x!$.]
\end{itemize}
\qex
\end{exercise}

\section*{Acknowledgements}
This work was supported in part by
NASA ATP award NAG5-10763 and by NSF grant AST-0205981,
and of course by this wonderful summer school.

\section*{Answers to multiple choice questions}

{\bf Question~\ref{choice}.}
The correct answer is D.
Whereas the other answers have to do with humanistics
(A \& B) or mathematics (C \& E),
D has to do with physics.
If density fluctuations in the universe were originally
generated as a superposition of many independent
random processes, then the resulting primordial
density field will be Gaussian.
This is a generic, albeit not universal,
prediction of inflation,
where density fluctuations are seeded by quantum
fluctuations of the inflaton field.
The prediction of Gaussianity
remains consistent with observations
(Komatsu et al 2003 \cite{K03}).

\skipp
{\bf Question~\ref{goodpk}.}
All of these are true except perhaps for E,
the truth of which depends on methodology.
For example, the Baumgart \& Fry (1991) \cite{BF91} method
described in \S\ref{trad}
is about as easy as could be,
but it is far from the best method.
Answers A--D are all important.
However, answer C is the most insightful,
because it gives the fundamental reason
-- statistical homogeneity --
for the power spectrum's superiority over the correlation function.
Answers A, B, \& D can all be construed as
consequences of the fundamental assumption of statistical homogeneity.
It is for essentially the same reason that CMB folk use the
spherical harmonic power spectrum $C_l$ rather
than the angular correlation function
to characterise fluctuations in the CMB.
The power spectrum $C_l$ is the covariance of spherical harmonics;
spherical harmonics are eigenmodes of the rotation operator;
CMB fluctuations are statistically rotation invariant;
hence the covariance matrix of CMB fluctuations must commute with
the rotation operator.
All else (the spherical harmonic analogue of answers A, B, \& D) follows.

\skipp
{\bf Question~\ref{priors}.}
This question generated some debate at the workshop.
My own ordering was A-F in the same order as written,
but many respectable people opined that B should come before A.
I might even agree with them.

\skipp
{\bf Question~\ref{Schpk}.}
The correct answer is C.
Answer A is {\em not\/} true,
and B is true only in the limit of vanishing shot noise.
Only the non-shot-noise-subtracted power spectra
can be expressed in the form of the Schwarz inequality~(\ref{Sch2}), as
\begin{equation}
\label{Schpkans}
  \left|
  {\left\langle \delta_g(\kvec) \delta_m(-\kvec) \right\rangle \over
  \left\langle | \delta_g(\kvec) |^2 \right\rangle^{1/2}
  \left\langle | \delta_m(\kvec) |^2 \right\rangle^{1/2}}
  \right|
  \leq 1
  \;.
\end{equation}
If you are concerned about the mixture of $\kvec$ and $-\kvec$
in equation~(\ref{Schpkans}),
then split $\delta(\kvec)$ into its real and imaginary parts
(which are uncorrelated, with equal variances),
and consider an estimator which is a sum
of the real and imaginary parts.
If you are concerned with the appearance of the vector wavevector $\kvec$
rather than its absolute value $k$,
then consider an estimator that is an arbitrarily weighted sum
of modes $\delta(\kvec)$ having the same wavenumber~$k$.


\begin{thebibliography}{99.}


\bibitem{BF91}
D. J. Baumgart, J. N. Fry:
Fourier spectra of three-dimensional data, ApJ \textbf{375}, 25 (1991)










\bibitem{FKP94}
H. A. Feldman, N. Kaiser, J. A. Peacock:
Power spectrum analysis of three-dimensional redshift surveys, ApJ \textbf{426}, 23--37 (1994)

\bibitem{F35}
R. A. Fisher:
The logic of inductive inference, J.\ Roy.\ Stat.\ Soc.\ \textbf{98}, 39--54 (1935)


\bibitem{KSL94}
K. B. Fisher, C. A. Scharf, O. Lahav:
A spherical harmonic approach to redshift distortion and a measurement of $\varOmega_0$ from the 1.2~Jy {\it IRAS\/} redshift survey,
MNRAS, 266, 219 (1994)

\bibitem{H97}
A. J. S. Hamilton:
Towards optimal measurement of power spectra - I. Minimum variance pair weighting and the Fisher matrix,
MNRAS \textbf{289}, 285 (1997)







\bibitem{HT95}
A. F. Heavens, A. N. Taylor:
A spherical harmonic analysis of redshift space, MNRAS \textbf{275}, 483--497 (1995)



\bibitem{KS67}
M. G. Kendall, A. Stuart:
\textit{The Advanced Theory of Statistics}
(Hafner Publishing, New York, 1967)

\bibitem{K03}
E. Komatsu et al (15 authors; WMAP collaboration)
First-Year Wilkinson Microwave Anisotropy Probe (WMAP) observations: tests of Gaussianity, ApJS \textbf{148}, 119 (2003)



\bibitem{P73}
P. J. E. Peebles:
Statistical analysis of catalogs of extragalactic objects. I. Theory, ApJ \textbf{185}, 413--440 (1973)






\bibitem{T97}
M. Tegmark:
How to measure CMB power spectra without losing information, Phys.\ Rev.\ D \textbf{55}, 5895 (1997)




\bibitem{TTH97}
M. Tegmark, A. Taylor, A. Heavens:
Karhunen-Loeve eigenvalue problems in cosmology: how should we tackle large data sets?,
ApJ \textbf{480}, 22 (1997)






\bibitem{YP69}
J. T. Yu, P. J. E. Peebles:
Superclusters of galaxies?, ApJ \textbf{158}, 103--113 (1969)

\end{thebibliography}
\end{document}